\documentstyle [preprint,aps,prb]{revtex}
\draft
\def\be{\begin{equation}}
\def\ee{\end{equation}}
\def\bea{\begin{eqnarray}}
\def\eea{\end{eqnarray}}

\begin{document}

\title{Reductive Perturbation Method, Multiple--Time Solutions and the KdV
Hierarchy}

\author{ R. A. Kraenkel$^1$, M. A. Manna$^2$, J. C. Montero$^1$, J. G.
Pereira$^1$ }
\vskip 0.5cm
\address{$^1$Instituto de F\'{\i}sica Te\'orica \\
         Universidade Estadual Paulista \\
	 Rua Pamplona 145 \\
	 01405-900 S\~ao Paulo -- Brazil
\vskip 1.0cm
$^2$Physique Math\'ematique et Th\'eorique, URA-CNRS 768 \\
         Universit\'e de Montpellier II \\
	 34095 Montpellier Cedex 05 -- France}
\maketitle
\begin{abstract}
We apply a multiple--time version of the reductive perturbation method to study
long waves as governed by the Boussinesq model equation. By requiring the
absence of secular producing terms in each order of the perturbative scheme, we
show that the solitary--wave of the Boussinesq equation can be written as a
solitary--wave satisfying simultaneously all equations of the KdV hierarchy,
each one in a different slow time variable. We also show that the conditions
for eliminating the secularities are such that they make the perturbation
theory compatible with the linear theory coming from the Boussinesq equation.
\end{abstract}
\vfill \eject

\section{Introduction}

As is well known, the intermediate long--wave Boussinesq model equation
\begin{equation}
u_{tt} - u_{xx} + u_{xxxx} - 3 (u^2)_{xx} = 0 \, ,
\label{1}
\ee
with $u(x,t)$ a one--dimensional field and with the subscripts denoting partial
differentiation, is completely integrable\cite{zakha} and has N--soliton
solutions. In particular, its solitary--wave solution is of the form\cite{soli}
\be
u = -2 k^2 \, {\rm sech}^2 \left[k \left( x - \sqrt{1 - 4 k^2} \; t \right)
\right] \, ,
\label{2}
\ee
where $k$ is the wavenumber.

To study the long waves of eq.(\ref{1}), we are going to consider a
perturbative scheme based on the reductive perturbation method of
Taniuti,\cite{taniu} modified by the introduction of an infinite number of slow
time-variables: $\tau_3$, $\tau_5$, $\tau_7$, etc. Then, as a consequence of a
natural compatibility condition, we have that any wave field satisfying the KdV
equation in the time $\tau_3$ must also satisfy all equations of the KdV
hierarchy,\cite{lax} each one in a different slow time variable. The main
reason for introducing these time variables, as we are going to see, is that
they allow for the construction of a perturbative scheme, valid for weak
nonlinear dispersive systems, which is free of solitary--wave related
secularities.

In this paper, by making use of the perturbative scheme with multiple slow
time--scales, we will show that the solitary--wave of the Boussinesq equation
may be written, in the slow variables, as a solitary--wave solution to the
whole set of equations of the KdV hierarchy, each one in a different slow time
variable. This result follows both, from the general long--wave perturbation
theory, and from the observation that the perturbative series truncates for a
solitary--wave solution to the KdV hierarchy equations, rendering thus an exact
solution for the Boussinesq equation. Furthermore, we will show that the
conditions for the elimination of the secular producing terms make the
perturbation theory compatible with the linear theory associated to the
Boussinesq equation.

\section{The Multiple Time Evolution Equations}

The long--wave limit is achieved by putting
\be
k = \epsilon \kappa \, ,
\label{lw}
\ee
where $\epsilon$ is a small parameter. Accordingly, we define a slow space
\be
\xi = \epsilon (x - t) \, ,
\label{10}
\ee
as well as an infinity of slow time coordinates:
\be
\tau_3 = \epsilon^3 t \quad ; \quad \tau_5 = \epsilon^5 t \quad ; \quad \tau_7
= \epsilon^7 t \quad ; \quad \dots
\quad .
\label{11}
\ee
Consequently, we have that
\be
\frac{\partial}{\partial x} = \epsilon \frac{\partial}{\partial \xi}
\label{12}
\ee
and
\be
\frac{\partial}{\partial t} = - \epsilon \frac{\partial}{\partial \xi} +
\epsilon^3 \frac{\partial}{\partial \tau_3} + \epsilon^5
\frac{\partial}{\partial \tau_5} + \epsilon^7 \frac{\partial}{\partial \tau_7}
+ \cdots \quad .
\label{13}
\ee
In addition, we make the expansion
\be
u = \epsilon^2 {\hat u} = \epsilon^2 \left(u_0 + \epsilon^2 u_2 + \epsilon^4
u_4 + \cdots \right) \, ,
\label{14}
\ee
and we suppose that $u_{2n} = u_{2n}(\xi , \tau_3 ,\tau_5 , ...),
n=0,1,2,\dots$, which corresponds to an extention in the sense of
Sandri.\cite{sandri} Substituting eqs.(\ref{12}), (\ref{13}) and (\ref{14})
into the Boussinesq equation (\ref{1}), the resulting expression, up to terms
of order $\epsilon^4$, is:
$$
\Big[- 2 \frac{\partial^2}{\partial \xi \partial \tau_3} + \frac{\partial^4}
{\partial \xi^4} + \epsilon^2 \left( \frac{\partial^2}{\partial
{\tau_3}^2} - 2 \frac{\partial^2}{\partial \xi \partial \tau_5} \right) +
\epsilon^4 \left(- 2 \frac{\partial^2}{\partial \xi \partial \tau_7} + 2
\frac{\partial^2}{\partial \tau_3 \partial \tau_5} \right) + \cdots \Big]
{\hat u}
$$
\be
- 3 \frac{\partial^2}{\partial {\xi}^2} \Big[ (u_0)^2 + 2 \epsilon^2 u_0 u_2
+ \epsilon^4 (2 u_0 u_4 + (u_2)^2) + \cdots \Big] = 0 \, .
\label{15}
\ee

We proceed now to an order--by--order analysis of this equation. At order
$\epsilon^0$, after an integration in $\xi$, we get
\be
\frac{\partial u_0}{\partial \tau_3} = \alpha_3 \left[ 6 u_0 \, \frac{\partial
u_0}{\partial \xi} - \frac{\partial^3 u_0}{\partial {\xi}^3} \right] = 0 \quad
; \quad \alpha_3 = - \frac{1}{2} \, ,
\label{17}
\ee
which is the KdV equation.

At order $\epsilon^2$, eq.(\ref{15}) yields
\be
\frac{\partial}{\partial \xi} \left[- 2 \frac{\partial u_2}{\partial \tau_3} -
6 \frac{\partial}{\partial \xi} (u_0 u_2) + \frac{\partial u_2}{\partial
{\xi}^3} \right] = 2 \frac{\partial^2 u_0}{\partial \xi \partial \tau_5} -
\frac{\partial^2 u_0}{\partial {\tau_3}^2} \, .
\label{18}
\ee
Using the KdV equation (\ref{17}) to express $\partial u_0 / \partial \tau_3$,
integrating once in $\xi$ and assuming a vanishing integration constant, we
obtain
\be
\frac{\partial u_2}{\partial \tau_3} + 3 \frac{\partial}{\partial \xi} (u_0
u_2) - \frac{1}{2} \frac{\partial^3 u_2}{\partial {\xi}^3} = - \frac{\partial
u_0}{\partial \tau_5} + \frac{1}{8} \frac{\partial^5 u_0}{\partial {\xi}^5}
- \frac{3}{2} u_0 \frac{\partial^3 u_0}{\partial {\xi}^3} - \frac{9}{4}
\frac{\partial u_0}{\partial \xi} \frac{\partial^2 u_0}{\partial {\xi}^2} +
\frac{9}{2} (u_0)^{2} \frac{\partial u_0}{\partial \xi} \; .
\label{19}
\ee
Equation (\ref{19}), as it stands, presents two problems. First, the evolution
of $u_0$ in the time $\tau_5$ is not known {\it a priori}. The second
problem is that the term $(\partial^5 u_{0} / \partial {\xi}^5 )$, as a source
term, is a secular producing term when $u_0$ is chosen to be a solitary--wave
solution of the KdV equation. In the next sections we will be dealing with
these two problems.

\section{The Korteweg--de Vries Hierarchy}

As we have seen, the field $u_0$ satisfies the KdV equation (\ref{17}) in the
time $\tau_3$. The evolution of the same field $u_0$ in any of the
higher--order times $\tau_{2n+1}$ can be obtained in the following
way.\cite{kmp} First, to have a well ordered perturbative scheme we impose that
each one of the equations for $u_{0\tau_{2n+1}}$ be $\epsilon$--independent
when passing from the slow $(u_0, \xi, \tau_{2n+1})$ to the laboratory
coordinates $(u, x, t)$. This step selects all possible terms to appear in
$u_{0\tau_{2n+1}}$. For instance, the evolution of $u_0$ in time $\tau_5$ is
restricted to be of the form
\be
u_{0\tau_5} = \alpha_5 u_{0(5\xi )} + \beta_5 u_0 u_{0\xi \xi \xi } + (\beta_5
+ \gamma_5) u_{0\xi }u_{0\xi \xi } + \delta_5 u_0^2 u_{0\xi } \, ,
\label{extra}
\ee
where $\alpha_5$, $\beta_5$, $\gamma_5$ and $\delta_5$ are unknown constants.
Then, by imposing the natural (in the multiple time formalism) compatibility
condition
\be
\Big( u_{0 \tau_3} \Big)_{\tau_{2n+1}} = \Big( u_{0\tau_{2n+1}} \Big)_{\tau_3}
\, ,
\label{compa}
\ee
with $u_{0\tau_3}$ given by eq.(\ref{17}), it is possible to determine any
$u_{0\tau_{2n+1}}$, {\it i.e.} to determine all constants appearing in
$u_{0\tau_{2n+1}}$. As it can be verified,\cite{kmp} the resulting equations
are those given by the KdV hierarchy. In particular, for $u_{0\tau_5}$ and
$u_{0\tau_7}$ we
obtain respectively
\be
u_{0\tau_5} = \alpha_5 \left[ u_{0(5\xi)} - 10 u_0 u_{0\xi\xi\xi} - 20 u_{0\xi}
u_{0\xi\xi} + 30 (u_{0})^2 u_{0\xi} \right] \, ,
\label{22}
\ee
and
\begin{eqnarray}
u_{0\tau_7} = \alpha_7 \big[&-& u_{0(7\xi)} + 14 u_0 u_{0(5\xi)} + 42 u_{0\xi}
u_{0(4\xi)} + 140 (u_0)^3 u_{0\xi} \nonumber \\
&+& 70 u_{0\xi\xi} u_{0\xi\xi\xi} - 280 u_0 u_{0\xi} u_{0\xi\xi} -
70 (u_{0\xi})^3 - 70 (u_{0})^2 u_{0\xi\xi\xi} \big] \, ,
\label{23}
\end{eqnarray}
where $\alpha_5$ and $\alpha_7$ are free parameters not determined by the
algebraic system originated from eq.(\ref{compa}). These free parameters are
related to different possible normalizations of the slow time variables.

\section{Higher Order Evolution Equations}

We return now to eq.(\ref{19}) for $u_2$. Substituting $u_{0\tau_5}$ from
eq.(\ref{22}), we obtain
\bea
u_{2\tau_3} + 3 (u_0 u_2)_{\xi} - \frac{1}{2} u_{2\xi\xi\xi} &=& \left[
\frac{1}{8} - \alpha_5 \right] u_{0(5\xi)} + \left[- \frac{3}{2} + 10 \alpha_5
\right] u_0 u_{0\xi\xi\xi} \nonumber \\
&+& \left[- \frac{9}{4} + 20 \alpha_5 \right] u_{0\xi} u_{0\xi\xi} +
\left[\frac{9}{2} - 30 \alpha_5 \right] (u_0)^2 u_{0\xi} \, .
\label{99}
\eea
We see thus that the secular--producing term $u_{0(5\xi)}$ can be eliminated if
we choose $\alpha_5=\frac{1}{8}$. In this case, Eq.(\ref{99}) becomes
\be
u_{2\tau_3} + 3 (u_0 u_2)_{\xi} - \frac{1}{2} u_{2\xi\xi\xi} = - \frac{1}{4}
\left[ - 3 (u_0)^2 u_{0\xi} + u_0 u_{0\xi\xi\xi} - u_{0\xi} u_{0\xi\xi} \right]
\, .
\label{26}
\ee

{}From this point on, we are going to consider some specific solutions to our
equations. First of all, we assume the solution of the KdV equation (\ref{17})
to be the solitary--wave solution
\be
u_0 = - 2 \kappa^2 {\rm sech^2} \left[\kappa \xi - 4 \alpha_3 \kappa^3 \tau_3 +
\theta \right] \, ,
\label{24}
\ee
where $\theta$ is a phase. However, we have just seen that $u_0$ must satisfy
also the higher order equation (\ref{22}) of the KdV hierarchy. Actually, as we
are going to see, to obtain a perturbative scheme free of secularities at any
higher order, we will assume that $u_0$ be a solitary--wave solution to all
equations of the KdV hierarchy, each one in a different slow--time variable.
Such a solution is given by
\be
u_0 = - 2 \kappa^2 {\rm sech^2} \left[\kappa \xi - 4 \alpha_3 \kappa^3 \tau_3
+ 16 \alpha_5 \kappa^5 \tau_5 - 64 \alpha_7 \kappa^7 \tau_7 + \cdots \right] \,
{}.
\label{25}
\ee
Using this solitary--wave solution, we see that the right--hand side of
eq.(\ref{26}) vanishes, leading to
\be
u_{2\tau_3} + 3 (u_0 u_2)_{\xi} - \frac{1}{2} u_{2\xi\xi\xi} = 0 \, ,
\label{27}
\ee
which is a homogeneous linearized KdV equation. We will assume for it the
trivial solution
\be
u_2 = 0 \, .
\label{28}
\ee

At order $\epsilon^4$, and already assuming that $u_2 =0$, eq.(\ref{15}) gives
\be
{u_4}_{\tau_3 \xi} + 3 (u_0 u_4)_{\xi \xi} - \frac{1}{2} {u_4}_{(4\xi)} = -
{u_0}_{\tau_7 \xi} + {u_0}_{\tau_3 \tau_5} \, .
\label{29}
\ee
Using equations (\ref{17}) and (\ref{22}) to express $u_{0\tau_3}$ and
$u_{0\tau_5}$ respectively, and integrating once in $\xi$, we obtain
\bea
u_{4\tau_3} &+& 3 (u_0 u_4)_{\xi} - \frac{1}{2} u_{4\xi\xi\xi} =
- u_{0\tau_7} + \frac{1}{16} u_{0(7\xi)} - u_0 u_{0(5\xi)} + \frac{45}{8}
(u_0)^2 u_{0\xi\xi\xi} \nonumber \\
&-& \frac{35}{8} u_{0\xi\xi} u_{0\xi\xi\xi} - \frac{5}{2} u_{0\xi} u_{0(4\xi)}
+ \frac{75}{4} u_0 u_{0\xi} u_{0\xi\xi} - \frac{45}{4} (u_0)^3 u_{0\xi} +
\frac{15}{4} (u_{0\xi})^3  \, .
\label{30}
\eea
The source term proportional $u_{0(7\xi)}$ is the only resonant, that is,
secular producing term to the solution $u_4$. Then, in the very same way we did
before, we first use the KdV hierarchy equation (\ref{23}) to express
$u_{0\tau_7}$. After we do that, we can then choose the free parameter
$\alpha_7$ in such a way to eliminate the resonant term from the right--hand
side of eq.(\ref{30}). This choice corresponds to $\alpha_7=-\frac{1}{16}$,
which brings eq.(\ref{30}) to the form
\bea
u_{4\tau_3} &+& 3 (u_0 u_4)_{\xi} - \frac{1}{2} u_{4\xi\xi\xi} =
\frac{1}{8} \big[ u_{0\xi} u_{0(4\xi)} - u_0 u_{0(5\xi)} \nonumber \\
&+& 10  u_0 u_{0\xi} u_{0\xi\xi} - 5 (u_{0\xi})^3 +
10 (u_0)^2 u_{0\xi\xi\xi} - 20 (u_0)^3 u_{0\xi} \big] \, .
\label{31}
\eea
Substituting again the solitary--wave solution (\ref{25}) for $u_0$, we can
easily
see that the nonhomogeneous term of eq.(\ref{31}) vanishes, leading to
\be
u_{4\tau_3} + 3 (u_0 u_4)_{\xi} - \frac{1}{2} u_{4\xi\xi\xi} = 0 \, .
\label{32}
\ee
And again, we take the trivial solution
\be
u_4 = 0 \, .
\label{33}
\ee
It is easy to see that this is a general result that will repeat at any higher
order: for $n \geq 1$, the evolution of $u_{2n}$ in the time $\tau_3$,
after using the KdV hierarchy equation to express $u_{0\tau_{2n+1}}$ and
substituting the solitary--wave solution (\ref{25}) for $u_0$, is given by a
homogeneous linearized KdV equation. Consequently, the solution $u_{2n} = 0$,
for $n \geq 1$, can be assumed for any higher order.

\section{Back to the Laboratory Coordinates}

Let us now take the solitary--wave solution to all equations of the KdV
hierarchy,
\be
u_0 = - 2 \kappa^2 {\rm sech^2} \left[\kappa \xi + 2 \kappa^3 \tau_3  + 2
\kappa^5 \tau_5 + 4 \kappa^7 \tau_7 + \cdots \right] \, ,
\label{35}
\ee
where we have already substituted the (not anymore) free parameters
$\alpha_{2n+1}$, and rewrite it in the laboratory coordinates. First, recall
that we have expanded $u$ according to eq.(\ref{14}). Thereafter, we have found
a particular solution in which $u_{2n} = 0$, for $n \geq 1$. Consequently,
expansion (\ref{14}) truncates, leading to an exact solution of the form
\be
u = \epsilon^{2} u_0 \, ,
\label{38}
\ee
with $u_0$ given by eq.(\ref{35}). Moreover, the slow variables
($\kappa,\xi,\tau_{2n+1}$) are related to the laboratory ones, ($k, x, t$),
respectively by eqs.(\ref{lw}), (\ref{10}) and (\ref{11}). Then,
in the laboratory coordinates, the exact solution (\ref{38}) is written as
\be
u = - 2 k^2 {\rm sech^2} \, k \left[x - \left( 1 - 2 k^2 - 2 k^4 - 4 k^6 -
\cdots
\right) t \right] \, .
\label{40}
\ee
Now, the series appearing inside the parenthesis can be summed:
\be
 1 - 2 k^2 - 2 k^4 - 4 k^6 - \cdots = \left(1 - 4 k^2 \right)^{1/2} \, \, .
\label{41}
\ee
Consequently, we get
\be
u = - 2 k^2 {\rm sech^2} \left[ k \left( x - \sqrt{1 - 4 k^2} \; t \right)
\right] \, ,
\label{42}
\ee
which is the well known solitary--wave solution of the Boussinesq
equation~(\ref{1}).

\section{Relation to the Dispersion Relation Expansion}

Let us take now the linear Boussinesq dispersion relation:
\be
\omega(k) = k \left(1 + k^2 \right)^{1/2} \, .
\label{dr}
\ee
Its long--wave ($k=\epsilon \kappa$) expansion is given by
\be
\omega(\kappa) = \epsilon \kappa + \alpha_3 \epsilon^3 \kappa^3 +
\alpha_5 \epsilon^5 \kappa^5 + \alpha_7 \epsilon^7 \kappa^7 +
\cdots \; ,
\label{dre}
\ee
where the coefficients $\alpha_{2n+1}$, except for $\alpha_3$ which arose
naturally in the KdV equation (\ref{17}), coincide exactly with those
necessary to eliminate the secular producing terms in each order of
the perturbative scheme. With this expansion, the solution of the associated
linear Boussinesq equation can be written as
\be
u = \exp i \left[\kappa \epsilon (x - t) + \alpha_3 \kappa^3
\epsilon^3 t + \alpha_5 \kappa^5 \epsilon^5 t + \alpha_7
\kappa^7 \epsilon^7 t + \cdots \right] \, .
\ee
Therefore, if we define from the very begining, as given by this solution, the
properly normalized slow time coordinates
\be
\tau_3 = \alpha_3 \epsilon^3 t \quad ; \quad \tau_5 =
\alpha_5 \epsilon^5 t \quad ; \quad \tau_7 =
\alpha_7 \epsilon^7 t \quad ; \quad \dots \; ,
\label{nst}
\ee
the resulting perturbative theory will be automatically free of
secularities.\cite{bous} Furthermore, the linear limit of the
perturbation theory will be compatible with the linear theory
coming directly from the Boussinesq equation (\ref{1}).

\section{Final Comments}

By applying a multiple time version of the reductive perturbation method to the
Boussinesq model equation, and by eliminating the solitary--wave related
secular producing terms through the use of the KdV hierarchy equations, we have
shown that the solitary--wave of the Boussinesq equation is given by a
solitary--wave satisfying, in the slow variables, all equations of the KdV
hierarchy. Accordingly, while the KdV solitary--wave depends only on one slow
time variable, namely $\tau_3$, the solitary--wave of the Boussinesq equation
can be thought of as depending on the infinite slow time variables.

The same results hold \cite{sww} for the shallow water wave (SWW) equation. In
other words, the solitary--wave of the SWW equation can also be written as a
solitary--wave satisfying simultaneously, in the slow variables, all equations
of the KdV hierarchy. It is important to remark that in both cases the
resulting secularity--free perturbation theory will be automatically compatible
with the corresponding linear theory.

A crucial point of the multiple time perturbative scheme is the return to the
laboratory coordinates, which implies in a renormalization of the
solitary--wave velocity.\cite{kota} In the case of the Boussinesq, as well as
of the SWW equation, this renormalization is such that the KdV hierarchy
solitary--wave is led to the corresponding Boussinesq or SWW solitary--waves.
However, when the original nonlinear dispersive system does not present an
exact solitary--wave solution, the series will not truncate. In this case, a
secularity--free expansion can still be obtained and the process of returning
to the laboratory coordinates can be made order--by--order at any higher order,
implying in a sucessive renormalization in the velocity of the solitary--wave,
which in this case is represented by the leading order term of the perturbative
series.\cite{kmp}

\section*{Acknowledgments}

The authors would like to thank J. L\'eon for useful discussions. They would
also like to thank CNPq-Brazil, FAPESP-Brazil and CNRS-France for financial
support.


\begin{thebibliography}{99}

\bibitem{zakha}
V. E. Zakharov and A. B. Shabat, Func. Anal. Appl. {\bf 8}, 226 (1974).
\bibitem{soli}
R. K. Dodd, J. C. Eilbeck, J. D. Gibbon and H. C. Morris, {\it Solitons and
Nonlinear Waves} (Academic, London, 1982).
\bibitem{taniu}
T. Taniuti, Suppl. Prog. Theor. Phys. {\bf 55}, 1 (1974).
\bibitem{lax}
P. D. Lax, Comm. Pure Appl. Math. {\bf 21}, 467 (1968).
\bibitem{sandri}
G. Sandri, Nuovo Cimento {\bf 36}, 67 (1965).
\bibitem{kmp}
R. A. Kraenkel, M. A. Manna and J. G. Pereira, J. Math. Phys. {\bf 36}, 307
(1995).
\bibitem{bous}
R. A. Kraenkel, M. A. Manna, J. C. Montero and J. G. Pereira, J. Math. Phys.
(1995), to appear.
\bibitem{sww}
R. A. Kraenkel, M. A. Manna, J. C. Montero and J. G. Pereira, {\it The
N--Soliton Dynamics of the Shallow Water Wave Equation and the Korteweg--de
Vries Hierarchy}, Preprint IFT-P.040/95.
\bibitem{kota}
Y. Kodama and T. Taniuti, J. Phys. Soc. Jpn, {\bf 45}, 298 (1978).
\end{thebibliography}
\end{document}